\documentclass[12pt]{article}
\usepackage{graphicx}
\usepackage{amssymb}
\usepackage{amsmath}
\numberwithin{equation}{section}

\def\cla{{\mathcal A}}

\def\sgn{{\rm sgn }}
\def\tr{{\bf tr }}
\newtheorem{Pa}{Paper}[section]
\newtheorem{Tm}[Pa]{{\bf Theorem}}

\newtheorem{Cy}[Pa]{{\bf Corollary}}
\newtheorem{Rk}[Pa]{{\bf Remark}}

\newtheorem{Ee}[Pa]{{\bf Example}}
\newtheorem{Dn}[Pa]{{\bf Definition}}
\newtheorem{Pn}[Pa]{{\bf Proposition}}
\newcommand{\E}{\mathrm{e}}

\title{The deviation factor and  divergences in quantum
electrodynamics, concrete examples}

\author{Lev Sakhnovich}

\date{}

\begin{document}

\maketitle


\noindent \emph{99 Cove ave. Milford, CT 06461, USA, }\\
\noindent E-mail: lsakhnovich@gmail.com\\



\begin{abstract}  We consider the divergences in quantum electrodynamics. Our approach is based on ideas from the theory of  generalized wave operators.
In particular, we use the concept of the deviation factor.
The deviation factor characterizes
the deviations of the initial and  final waves from the free waves.
The approach is demonstrated on important examples.
\end{abstract}
\section{Introduction} \label{intro}
  1.
 In  QED,  the higher order approximations of the
elements of the scattering matrix contain integrals which diverge.
We suppose that these divergences appear as a result of the  representation of the scattering matrix in the form of
the power series in a small parameter e.

In Section 2, we try to answer J.R. Oppenheimer's question \cite{Opp}: 
``Can the procedure be freed from the expansion in e and carried out rigorously?" \\
The answer is positive, if the method of the generalized wave operators (GW-method) \cite{Dol, Sakh2, Sakh3, Sakh8, BuM, Pear} is applicable.

For the convenience of the readers, we introduce  the notions of the generalized wave operators
$W_{\pm}(A,A_0)$ and generalized scattering operator $S(A,A_0)$ in the Appendix (see Definitions 
\ref{Definition 7.1} and \ref{Definition 7.2}).
Here $A_0$ is a non-perturbed operator and 
$A$ is a perturbed operator. The generalized wave operators and generalized scattering operator are defined with the help of the  deviation factor $W_{0}(t)$ (see Definition 
\ref{Definition 7.1}). The deviation factors   characterize the deviations of the initial and final  waves
from the free waves. It is important that the generalized scattering operator has many properties of the classical scattering operator (see e.g.  \eqref{7.5} and \eqref{7.9}). 

We stress that the GW-method   does not use the power series. First we prove that the limit in \eqref{7.1} exists and then construct the generalized scattering operator using  
relation \eqref{7.4}.
In Section 2, we show that in the simple but important cases of the radial  Schr\"{o}dinger equation (Example 2.1) and of the radial Dirac equation (Example 2.2) with Coulomb-type potentials the GW-method  is applicable. Hence  in these cases the corresponding scattering theory is rigorous.

The GW-method  cannot be used directly in the theory of the Feynman integrals because the operators $A$ and $A_{0}$ do not exist in this theory.
However, Examples 2.1, 2.2 and the GW-method serve as a model for working with Feynman integrals. In order to find the connection between the GW-method  and Feynman integrals we consider again Example 2.1 and
 write the power series \eqref{2.16}:
\begin{equation}S(t,\tau)=I-S_{1}(t,\tau)e-...,\label{1.1}\end{equation}
 Then, using deviation factor $W_{0}(t)$  we write  the new power series \eqref{2.26}: 
 \begin{equation}S^{R}(t,\tau):=W_{0}(t)S(t,\tau)W_{0}^{-1}(\tau)=I-S_{1}^{R}(t,\tau)e-...,\label{1.2}\end{equation}
 The expression  $\lim[S(t,\tau)]$ when $t{\to}+\infty$ and $\tau{\to}-\infty$ does not exist but
the expression  $\lim[S^{R}(t,\tau)]$ when $t{\to}+\infty$ and $\tau{\to}-\infty$ exists.

Thus, we  answer the mentioned above J.R. Oppenheimer's question in the following way:
{\it We replace the original
series \eqref{2.16} by the new series \eqref{2.26}, which allows us to
make our  arguments rigorous.}

We use the  outlined scheme   by  the study of  the following cases: infrared problems, ultraviolet problems, Feynman integrals,
dimensional regularization.  For each of these cases we introduce the appropriate definition of the
deviation factor. For simplicity, we consider only the lowest order of approximation.
 We do not remove the divergences: our aim is to prove
that they are absent.

We note that the classical theory of divergences just removes the divergent integrals. In our approach these divergent integrals   obtain a
physical  meaning.  That is, using these integrals we construct the  deviation factors.
The notion of the deviation factor was introduced in \cite{Sakh5, Sakh4}.

 \emph{Summing up, we did not only derive the exact results in the theory of the divergences, but obtained also some new facts regarding the behaviour of the
 system, see \eqref{7.8}. It seems that these facts may be  checked experimentally as well.}
 
 Interesting concrete examples illustrating our method are given in Section~4.
 In particular, in Example 4.1 we deal  with the electron collision and in Example 4.5  we deal with the second order photon self-energy part.
An important method for removing divergences is based on the transition to the dimension $4-i\varepsilon,\quad \varepsilon{\to}+0$ (see \cite{COL, POK}).
In Section 5, we  show that our approach
 is  useful for such problems too.

 The  theory of generalized wave operators was used in an important paper by P.P. Kulish and L.D. Faddeev \cite{KF} for solving infrared divergence problems. Our approach to the divergence problems is essentially different from the
 Kulish--Faddeev approach. In particular, the scattering operator  commutes in our case with the unperturbed operator (see Section 2),
 which does not happen
in the Kulish--Faddeev theory.
\section{Coulomb-type infrared divergence}
\subsection{Nonstationary Schr\"{o}dinger equation}
\begin{Ee}\label{Example 2.1} Consider the nonstationary Schr\"{o}dinger equation
\begin{equation}i\frac{d\psi}{dt}= {\mathcal{L}}\psi, \label{2.0}\end{equation}
where ${\mathcal{L}}$ is the 
radial  Schr\"{o}dinger operator:
\begin{equation}\mathcal{L}f=-\frac{d^{2}}{dr^{2}}f+[\frac{\ell(\ell+1)}{r^{2}}
-\frac{2ez}{r}+eq(r)]f,\quad z=\overline{z} \label{2.1}\end{equation}
with Coulomb-type potential $2e(z/r)-eq(r)$, where $e$ is the elementary charge and $q(r)=\overline{q(r)}$ satisfies the conditions 
\begin{equation} \int_{0}^{1}|q^{2}|r^{2}dr<\infty;\quad  \int_{1}^{\infty}|q^{n}(r)|dr<\infty \quad  (n=1,2).\label{2.2}\end{equation}
\end{Ee}
Introduce  the operator $\mathcal{L}_{0}$:
\begin{equation}\mathcal{L}_{0}f=-\frac{d^{2}}{dr^{2}}f,\label{2.3}\end{equation}
and the boundary condition
\begin{equation}f(0)=0.\label{2.4}\end{equation}
The  statement below was proved in our paper \cite[p. 210]{Sakh2}.
\begin{Tm}\label{Theorem 2.2} Suppose that $q(r)$ satisfies relations \eqref{2.2}. Then, the generalized wave operators $W_{\pm}(\mathcal{L},\mathcal{L}_{0})$  and
the generalized scattering operator $S(\mathcal{L},\mathcal{L}_{0})$ exist. The corresponding deviation factor has the form
\begin{equation}W_{0}(t)=|t|^{-i \, \sgn(t) \, (ez)/\sqrt{\mathcal{L}_0}}.\label{2.5}\end{equation}
\end{Tm}
Thus, if conditions \eqref{2.2} are fulfilled, then relations \eqref{7.7}-\eqref{7.9}, where $A={\mathcal{L}}$ and $A_0={\mathcal{L}_0}$, are valid.
In particular, we have
\begin{equation}S(\mathcal{L},\mathcal{L}_{0})\psi_{-}=\psi_{+}.\label{2.6}\end{equation}
\begin{Rk}\label{Remark 2.3} We do not use the power series and obtain the rigorous result \eqref{2.6} in terms
of the generalized wave and scattering operators.\end{Rk}

Now, we shall consider the deviation factor given in \eqref{2.5} in greater detail {\it in order to explain the role of the deviation factor and its
connection with divergence problems}. 
We assume that  $q(r)$ is the superposition of the Yukawa potentials, that is, $q(r)$ has the form
\begin{equation}q(r)=\frac{1}{r}\int_{m}^{\infty}\exp(-r\beta)d\rho(\beta),\quad \int_{m}^{\infty}d|\rho(\beta)|<\infty,\quad m>0.\label{2.7}\end{equation}
It is easy to see that the introduced potential $q(r)$ satisfies the conditions \eqref{2.2}.
Using Levy results \cite{Le}, we write down the operators $\mathcal{L}_{0}$  and $\mathcal{L}$ in the momentum representation:
\begin{align}&\mathcal{L}_{0}f=k^{2}f,\quad f{\in}L^{2}(0,\infty),\label{2.8}
\\ &
\mathcal{L}f=k^{2}f+e\int_{0}^{\infty}f(p)pkR_{\ell}(k,p)dp, \label{2.9}\end{align}
where
\begin{equation}R_{\ell}(k,p)=-\frac{2z}{{\pi}kp}Q_{\ell}(\frac{k^2+p^2}{2pk})+
\frac{2}{\pi}\int_{m}^{\infty}\int_{-1}^{1}P_{\ell}(x)\frac{dxd{\rho}(\beta)}{k^2+p^2-2pkx+\beta^{2}}.
\label{2.10}\end{equation}
Here, $P_{\ell}(x)$ are Legendre polynomials and $Q_{\ell}(x)$ are Legendre polynomials of the second kind.

According to \eqref{2.5}, the corresponding deviation factor (in the momentum representation) has the form .
\begin{equation}W_{0}(t)f(k)=|t|^{-i(ez/k){\sgn}(t)}f(k),\quad f(k){\in}L^{2}(0,\infty).\label{2.11}\end{equation}
Then, the generalized scattering operator $S(\mathcal{L},\mathcal{L}_{0})$ exists and can be written in the form
\begin{equation}S(\mathcal{L},\mathcal{L}_{0})={\lim}W_{0}(t,k)S(t,\tau)W_{0}^{-1}(\tau,k)
\quad (t{\to}+\infty,\quad \tau{\to}-\infty),\label{2.12}\end{equation}
where
\begin{equation} S(t,\tau)=\exp(it\mathcal{L}_{0})\exp(-it\mathcal{L})\exp(i\tau\mathcal{L})\exp(-i\tau\mathcal{L}_{0}).
\label{2.13}\end{equation}
If $q(r)\equiv 0$, the generalized scattering operator can be evaluated explicitly.                            
In the momentum representation it reduces to multiplication by the function \cite[p. 215]{Sakh2}:
\begin{equation}S(\mathcal{L},\mathcal{L}_{0})f(k)=[\Gamma(\ell+1-i\frac{ze}{k})/\Gamma(\ell+1+i\frac{ze}{k})](2k)^{4ize/k}f(k),\label{2.14}
\end{equation}
where $\Gamma(z)$ is Euler gamma function. Taking into account \eqref{2.1} and \eqref{2.3}, we have the following assertion.
\begin{Cy}\label{Corollary 2.4}If $q(r) \equiv 0$ $($the case of pure  Coulomb potential$)$ and $e$ is small, then the corresponding generalized scattering operator can be represented in the form of the convergent power series of $e$:
\begin{equation}S(\mathcal{L},\mathcal{L}_{0})=1-2[1-2\log(2k)]i(z/k)e+... \label{2.15}\end{equation}\end{Cy}
Next we shall consider the case when $q(r)$ admits representation \eqref{2.7}.
 Let us write $S(t,\tau)$  in the series form \cite{AB}:
\begin{equation} S(t,\tau)=I-\sum_{p=1}^{\infty}S_{p}(t,\tau)e^{p},\label{2.16}\end{equation}
where
\begin{align}& S_{p}(t,\tau)=(-i)^{p}\int_{\tau}^{t}\int_{\tau}^{t_{1}}...
\int_{\tau}^{t_{p-1}}V(t_{1})V(t_{2})...V(t_{p})dt_{p}...dt_{1},\label{2.17}
\\ &
V(t)=\exp(it\mathcal{L}_{0})\mathcal{L}_{1}\exp(-it\mathcal{L}_{0}),
\quad \mathcal{L}=\mathcal{L}_{0}+e\mathcal{L}_{1}.\label{2.18}\end{align}
We shall calculate the first term $S_{1}(t,\tau)$ of the series \eqref{2.16}:
\begin{equation}S_{1}(t,\tau)=-i\int_{\tau}^{t}V(t_{1})dt_{1}.\label{2.19}\end{equation}
Taking into account \eqref{2.8}, \eqref{2.9} and \eqref{2.17}, \eqref{2.18} we obtain
\begin{equation}V(t)f(p)=\int_{0}^{\infty}pkR_{\ell}(k,p){\E}^{it(k^2-p^2)}f(p)dp,\label{2.20}\end{equation}
where $\E$ is the base of the natural logarithm.
It follows from \eqref{2.14} that
\begin{align}&\frac{2iz}{\pi}\int_{\tau}^{t}\int_{0}^{\infty}Q_{\ell}(\frac{k^2+p^2}{2kp}){\E}^{it_{1}(k^2-p^2)}f(p)dpdt_{1}\nonumber
\\ &
{\sim}f(k)[\frac{iz}{k}\log|t\tau|-2i\frac{\Gamma^{\prime}(\ell+1)}{\Gamma(\ell+1)}+
i\frac{4z}{k}\log{2k}] \quad (t{\to}+\infty,\, \tau{\to}-\infty).\label{2.21}\end{align}
For continuous kernels $T(k,p)$ the formula
\begin{equation}\int_{0}^{\infty}f(p)T(k,p)\frac{1}{k^2-p^2}[{\E}^{i(k^2-p^2)t}-{\E}^{i(k^2-p^2)\tau}]dp\, {\sim}\,
f(k)i{\pi}T(k,k)/k\label{2.22}\end{equation} is valid.
Here  $t{\to}+\infty,\,\tau{\to}-\infty.$ Relations \eqref{2.10} and \eqref{2.19}-\eqref{2.22} yield (see \cite{Sakh3})
that
\begin{equation}S_{1}(t,\tau)f(k) \,{\sim}\, f(k)[\frac{iz}{k}\log|t\tau|+\hat{S}_{1}(k,\ell)],\label{2.23}\end{equation}
where
\begin{align} \hat{S}_{1}(k,\ell)=&-2i\frac{z\Gamma^{\prime}(\ell+1)}{k\Gamma(\ell+1)}+i\frac{4z}{k}\log{2k}\nonumber
\\ &
-\frac{2i}{k}
\int_{m}^{\infty}\int_{-1}^{1}P_{\ell}(x)\frac{k}{2k^{2}(1-x)+\beta^{2}}dxd\rho(\beta).
\label{2.24}\end{align}Formula \eqref{2.23} shows that  $S_{1}(t,\tau){\to}\infty$  when $t{\to}+\infty,\quad \tau{\to}-\infty.$ The deviation factor $W_{0}(t)$ (see \eqref{2.11}) can be represented in the form of the convergent power series of $e$:
\begin{equation}W_{0}(t)=1-i[(z/k)(\sgn(t))\log|t|]e+...\label{2.25}\end{equation}
Using \eqref{2.16} and \eqref{2.25} we write
\begin{equation}W_{0}(t)S(t,\tau)W_{0}^{-1}(\tau)= I-\sum_{p=1}^{\infty}S_{p}^{R}(t,\tau)e^{p}.\label{2.26}\end{equation}
It follows from \eqref{2.21}, \eqref{2.22}, \eqref{2.24} and \eqref{2.26} that
\begin{equation}S_{1}^{R}(t,\tau)\,{\sim}\, \hat{S}_{1}(k,\ell),\quad t{\to}+\infty,\quad \tau{\to}-\infty.\label{2.27}\end{equation}
Hence, the deviation $W_{0}(t)$ factor has a property described in proposition below.
\begin{Pn}\label{Proposition 2.5} The first term $S_{1}^{R}(t,\tau)$ of the series \eqref{2.26} for the generalized scattering operator is regular,
i.e. it has a finite limit, when  $t{\to}+\infty,\quad \tau{\to}-\infty.$\end{Pn}
\begin{Rk}\label{Remark 2.6} Using Corollary 2.4 we can prove that  all terms  $S_{p}^{R}(t,\tau)$ of the series \eqref{2.26} for the generalized scattering operator 
are regular.\end{Rk}
\subsection{Nonstationary Dirac equation}
\begin{Ee}\label{Example 2.7} 
Consider the nonstationary Dirac equation
\begin{equation}i\frac{d\psi}{dt}= {\mathcal{L}}\psi, \label{2.0'}\end{equation}
where ${\mathcal{L}}$ is the radial Dirac operator. This operator acts in the space $L_{2}^{2}(0,\infty)$ and has the form
\begin{equation}\mathcal{L}F=[i \sigma_2\frac{d}{dr}+\sigma_{1}\frac{k}{r}+v(r)-m\sigma_{3}]F,\label{2.31}\end{equation}
where $F={\mathrm{col}}[f,g],\,\, v(r)=\frac{\cla e}{r}-eq(r)$, $\, m>0,\,\, k=\overline{k}$, and
\begin{equation} \cla =\overline{\cla }{\ne}0,\quad |k|>e|\cla |.\label{2.30}\end{equation}
Recall that the Pauli matrices $\sigma_i$ are given by the equalities
\begin{equation}\sigma_1=\left(
                           \begin{array}{cc}
                             0 & 1 \\
                             1 & 0 \\
                           \end{array}
                         \right),\quad
\sigma_2=\left(
                           \begin{array}{cc}
                             0 & -i \\
                             i & 0 \\
                           \end{array}
                         \right),\quad
\sigma_3=\left(
                           \begin{array}{cc}
                             1 & 0 \\
                             0 & -1 \\
                           \end{array}
                         \right).\label{4.2}\end{equation}
We note that the corresponding  radial Dirac  equation has the form
\begin{align}&
\left( \frac{d}{dr}+\frac{k}{r}\right) f-\big(\lambda+m+\frac{e\cla }{r}-eq(r)\big) g=0,
\label{2.28}
\\ &
\left( \frac{d}{dr}-\frac{k}{r}\right) g+\big(\lambda-m+\frac{e\cla }{r}-eq(r)\big) f=0.
\label{2.29}\end{align}
where $\lambda = \overline{\lambda}$ and  $e$ is the elementary charge.
\end{Ee}
Let us introduce the non-perturbed operator
\begin{equation}\mathcal{L}_{0}F=[i \sigma_2\frac{d}{dr}-m\sigma_{3}]F,\label{2.33}\end{equation}
Assume that the following relations are valid:
\begin{equation}\int_{0}^{a}|rq(r)|dr+\int_{a}^{\infty}|q(r)|dr<\infty,\quad q(r)=\overline{q(r)}.\label{2.34}\end{equation}
The following statement is proved in our papers \cite{Sakh1, Sakh5}.
\begin{Tm}\label{Theorem 2.8}Suppose that $q(r)$ satisfies  \eqref{2.34}. Then, the generalized wave operators $W_{\pm}(\mathcal{L},\mathcal{L}_{0})$  and
the generalized scattering operator $S(\mathcal{L},\mathcal{L}_{0})$ exist. The corresponding deviation factor has the form
\begin{equation}W_{0}(t)=|t|^{i\sgn(t)e \cla\phi(\mathcal{L}_{0})},
\label{2.35}\end{equation}
where
\begin{equation}\phi(\mathcal{L}_{0})=\mathcal{L}_{0}(\mathcal{L}_{0}^2-m^2)^{-1/2}.\label{2.36}\end{equation}
\end{Tm}
Thus, if conditions \eqref{2.34} are fulfilled, then the relations \eqref{7.7}-\eqref{7.9} (with $A=\mathcal{L}$ and $A_0=\mathcal{L}_0$ given by \eqref{2.31}
and \eqref{2.33}) are valid for the solutions $\psi$ of the nonstationary Dirac equation \eqref{2.0'}.

\begin{Rk}\label{Remark 2.9}  In the Examples \ref{Example 2.1} and \ref{Example 2.7}, we considered radial Schr\"odinger and Dirac equations
with Coulomb-type potentials.  When one uses the classical approach, divergence problems appear in such cases. However, we showed that
the GW-method allows to derive a rigorous scattering theory without divergences.
\end{Rk}
\subsection{Some further scattering problems}
In a number of scattering problems, the  perturbed operators $A$ and non-perturbed operators $A_0$ do not exist.
Some information is known only about the scattering operator. In these cases the wave operators (usual and generalized) also do not exist. But deviation factor can be introduced in these cases too. Thus,  one may use here the {\it deviation factor method} instead of the GW-method. We shall explain the situation using
the following model example.
\begin{Ee}\label{Example 2.10}
 Let the   element $d(q)$ of the  scattering operator
 be represented in the form $d(q)=\lim{d(q,t,\tau)},\quad t{\to}+\infty,\,\tau{\to}-\infty$, and
\begin{equation}d(q,t,\tau)= 1+ea_{1}(q,t,\tau)+o(e),
\label{2.37}\end{equation}
where $e$ is the elementary charge.\end{Ee}
If
\begin{equation}a_{1}(q,t,\tau)=i\phi(q)\ln{|t\tau|}+O(1) \quad {\mathrm{for}} \quad 
t{\to}+\infty,\,\, \tau{\to}-\infty, \label{2.38}\end{equation}
where $\phi(q)=\overline{\phi(q)}$, then the right-hand side of \eqref{2.37} diverges logarithmically.
Relation \eqref{2.38} holds in the case of the Coulomb-type potentials (see Examples \ref{Example 2.1} and \ref{Example 2.7}).
\begin{Dn}\label{Definition 2.11}We say that the coefficient $a_{1}(q,t,\tau)$ has Coulomb-type divergence
if \eqref{2.38} is valid $($and $\phi(q)=\overline{\phi(q)})$.\end{Dn}
We note that the Coulomb-interaction is the typical example of infrared catastrophe \cite{CD, Ku, KF}.
Introduce $\tilde d$ by the formula
\begin{equation}\tilde{d}(q,t,\tau)=
|t\tau|^{-ie\phi(q)}d(q,t,\tau).
\label{2.39}\end{equation}
Using  \eqref{2.37} and \eqref{2.39} we have
\begin{equation}\tilde{d}(q,t,\tau)=1+e[a_{1}(q,t,\tau)-i\phi(q)\ln|t\tau|]+...
\label{2.40}\end{equation}
The following assertion is valid.
\begin{Pn}\label{Proposition 2.12} The first term
\begin{equation}\tilde{a}_{1}(q,t,\tau)=a_{1}(q,t,\tau)-i\phi(q)\ln|t\tau| \label{2.41}\end{equation}
of power series \eqref{2.40} is regular, that is, it  converges when $t{\to}+\infty$ and $\tau{\to}-\infty$.\end{Pn}
Now, we  introduce the deviation factor $W_{0}(t)$ for the case \eqref{2.37}.
\begin{Dn}\label{Definition 2.13} The function $W_{0}(t)$ is the deviation factor
for  the case \eqref{2.37} if the following conditions are fulfilled:\\
1) $|W_{0}(t)|=1$ and  $W_{0}(t)W_{0}(t+\tau)^{-1}{\to}1,\quad t{\to}\pm\infty.$\\
2) The first term
$\tilde{a}_{1}(q,t,\tau)$
of power series
\begin{equation}\tilde{d}(q,t,\tau)=W_{0}(t)d(q,t,\tau)W_{0}^{-1}(\tau)=1+\tilde{a}_{1}(q,t,\tau)e+o(e)
\label{2.42}\end{equation}
is regular, that is, it  converges when $t{\to}+\infty$ and $\tau{\to}-\infty$.
\end{Dn}
Condition 1) from Definition \ref{Definition 2.13} is  prompted by the theory of generalized wave operators (see Appendix).
Below, we formulate a corollary of Proposition \ref{Proposition 2.12}.
\begin{Cy}\label{Corollary 2.14} If \eqref{2.38} holds, then
the corresponding deviation factor $W_{0}(t)$  has the form
\begin{equation}W_{0}(t)=|t|^{-ie\sgn(t)\phi(q)}.\label{2.43}\end{equation}\end{Cy}
\section{Ultraviolet divergence}
 Let the  element $d(q)$ of the  scattering operator   be given by \\ $d(q)=\lim_{L\to \infty}{d(q,L)}$, where
\begin{equation}d(q,L)=1+e^{2}a_{2}(q,L)+o(e^2),
\label{3.1}\end{equation}
and $e$ is the elementary charge.
We assume that
\begin{equation}
a_{2}(q)=\lim_{L{\to}\infty}\int_{\Omega}F(P,Q)d^{4}P,\label{3.2}\end{equation}
where $P=[-ip_0,p_1,p_2,p_3]$, $Q=[-iq_0,q_1,q_2,q_3]$ and  the invariant region of integration
$\Omega$ is the four dimensional sphere with the radius $L$.

We shall investigate the cases when the limit on the right-hand side of \eqref{3.2}
does not exist. Then we have the ultraviolet divergences. It is known that the corresponding divergences can be removed by mass and charge 
renormalization \cite{AB}. F.J. Dyson \cite{Dy}  stressed that it is important ``to prove the convergence in the frame of the theory". We shall do it for a  wide
class of examples.
\begin{Ee}\label{Example 3.1} Let us consider the case when
 \begin{equation}a_{2}(q,L)=\int_{\Omega}F(p,q)d^{4}p=i[\phi(q)\ln{L}+\psi(q)+O(1/L)],
\,\, L{\to}+\infty.\label{3.3}\end{equation}
where $\phi(q)=\overline{\phi(q)}$. \end{Ee}
We see that  the integral in \eqref{3.3} diverges logarithmically. 

\emph{Now, we
 use a new representation of $d(q,L)$}: 
\begin{equation}d(q,L)=L^{{ie}^{2}\phi(q)}\tilde{d}(q,L).
\label{3.4}\end{equation}
Taking into account \eqref{3.1} and \eqref{3.4} we obtain a  representation of $\tilde{d}(q,L)$:
\begin{equation}\tilde{d}(q,L)=1+e^{2}
[a_{2}(q,L)-i\phi(q)\ln{L}]+o(e^2).
\label{3.5}\end{equation}
The following assertion is valid:
\begin{Pn}\label{Proposition 3.2} The second  term
 $\tilde{a}_{2}(q,L)=a_{2}(q,L)-i\phi(q)\ln{L}$
of relation \eqref{3.5} is regular, that is, it  converges when $L{\to}+\infty$.  \end{Pn}

Next we  introduce the deviation factor $W_{0}(L)$ for the case \eqref{3.1}.
\begin{Dn}\label{Definition 3.3} The function $W_{0}(L)$ is the deviation factor
for  the case \eqref{3.1} if the following conditions are fulfilled:\\
1) $|W_{0}(L)|=1$ and  $W_{0}(L)W_{0}(L+\ell)^{-1}{\to}1,\quad L{\to}\infty.$\\
2) The second  term
$\tilde{a}_{2}(q,L)$
of relation
\begin{equation}\tilde{d}(q,t,\tau)=W_{0}^{-1}(L)d(q,L)=1+\tilde{a}_{2}(q,L)e^{2}+o(e^2)
\label{3.6}\end{equation}
is regular, that is, it  converges when $L{\to}+\infty.$
\end{Dn}
We obtain a corollary of Proposition \ref{Proposition 3.2}.
\begin{Cy}\label{Corollary 3.4} If relation \eqref{3.3} holds, then
the corresponding deviation factor $W_{0}(L)$  has the form
\begin{equation}W_{0}(L)=L^{ie^{2}\phi(q)}.\label{3.7}\end{equation}\end{Cy}
\begin{Rk}\label{Remark 3.5}The element $d(q)$ of the  scattering operator  may be a matrix as well. In this case, the
coefficient $a_{2}(q,L)$  in \eqref{3.1} and the corresponding
deviation factor $W_{0}(L)$ are matrices. The Definition \ref{Definition 3.3} is valid in the matrix case too but condition 1)
takes in the matrix case the form:\\
1) $W_{0}(L)$ is a unitary matrix and  $W_{0}(L)W_{0}(L+\ell)^{-1}{\to}I,\quad L{\to}\infty.$
\end{Rk}
\begin{Rk}\label{Remark 3.6} Deviation factors are not uniquely defined. If $W_{0}(L)$ is the deviation factor and $C$ is a unitary operator,
then $CW_{0}(L)$ is  the deviation factor too.
The choice of multipliers $C$ depends on the particular physical problem
under consideration.\end{Rk}
\section{Concrete examples, Feynman integrals}
Let us illustrate the deviation factor  method  by further concrete examples. We will show that in these examples
the conditions of Definition  \ref{Definition 3.3} are fulfilled.
Introduce the   Dirac  matrices $\alpha_{k}$ and $\gamma_{k}$:
\begin{equation}\alpha_k=\left(
                           \begin{array}{cc}
                             0 & \sigma_k \\
                             \sigma_k & 0 \\
                           \end{array}
                         \right) \quad (k=1,2,3), \quad \gamma_{j}=\beta\alpha_{j}\quad (j=1,2,3),\quad \gamma_4=\beta,
                         \label{4.1}\end{equation}
where $\sigma_k$ are Pauli matrices (see \eqref{4.2}), $\beta=\left(
                         \begin{array}{cc}
                            I_2 & 0 \\
                           0 & -I_2 \\
                         \end{array}
                       \right)$, and $I_2$ is the $2\times 2$ identity matrix.
The matrices  $\hat{p}$ and  $\hat{k}$  are defined by the equalities
\begin{equation}
\hat{p}=\sum_{\mu}p_{\mu}\gamma_{\mu}, \quad \hat{k}=\sum_{\mu}k_{\mu}\gamma_{\mu}.
\label{4.4}\end{equation}
\begin{Ee}\label{Example 4.1} Let the element $d(p)$ of the scattering operator be given by the equality
 $d(p)=\lim_{L{\to}\infty}{d(p,L)}$, where
\begin{equation}d(p,L)=1+e^{2}a_{2}(p,L)+o(e^2),
\label{4.5}\end{equation}
$d(p,L)$ is a $4{\times}4$ matrix and the second term $a_{2}(p,L)$ has the form
\begin{equation}a_{2}(p,L)=J_{\mu}(L)=\frac{1}{(2\pi)^4}\int_{Q}\gamma_{\mu}\frac{i(\hat{p}-\hat{k})-m}{(p-k)^2+m^2}
\gamma_{\mu}\frac{dk}{k^2}.\label{4.6}\end{equation}
Here, $k=[k_1,k_2,k_3,k_4]$,  $p=[p_1,p_2,p_3,p_4],\, \,p^2=p_{1}^2+p_{2}^2+p_{3}^2+p_{4}^2,$ $Q$ is
the four dimensional sphere with the radius $L$, and $J_{\mu}(L)$ is the Feynman integral.\end{Ee}
We note that the integral $J_{\mu}(L)$ plays an important role in electron collision problems \cite{AB}.
Using the standard Feynman parameters
\begin{equation}\frac{1}{ab}=\int_{0}^{1}\frac{du}{[au+b(1-u)]^2}, \label{4.7}\end{equation}
we represent $J_{\mu}(L)$ in the form
\begin{equation}J_{\mu}(L)=\frac{1}{(2\pi)^4}\int_{0}^{1}\int_{Q}\gamma_{\mu}\frac{i(\hat{p}-\hat{k})-m}{(k^2-2pku+\ell)^2}\gamma_{\mu}dkdu,
\label{4.8}\end{equation}
where $\ell=(p^2+m^2)u.$
Let us write two well- known formulas \cite{AB}:
\begin{equation}\int_{Q}\frac{dk}{(k^2-2pk+\ell)^2}dk=i\pi^{2}(\ln\frac{L^2}{\ell-p^{2}}-1)+o(1),
\quad L{\to}\infty,\label{4.9}\end{equation}
\begin{equation}\int_{Q}\frac{k_{\nu}dk}{(k^2-2pk+\ell)^2}=i\pi^{2}p_{\nu}(\ln\frac{L^2}{\ell-p^{2}}-3/2)+o(1),
\quad L{\to}\infty.\label{4.10}\end{equation}
It follows from formulas \eqref{4.9} and \eqref{4.10} that
\begin{equation}\int_{Q}\frac{i(p_{\nu}-k_{\nu})}{(k^2-2pku+\ell)^2}dk=-\frac{\pi^2}{2}p_{\nu}
+o(1),
\quad L{\to}\infty.\label{4.11}\end{equation}
Using equalities
\begin{equation}\gamma_{\mu}^{2}=-I_{4} \quad (\mu=1,2,3),\quad \gamma_{4}^{2}=I_{4}, \label{4.12}\end{equation}
and relations  \eqref{4.8}, \eqref{4.9} and  \eqref{4.11}, we obtain the equalities
 \begin{equation}J_{\mu}(L)=\frac{1}{(2\pi)^4}\big[m\, i \, \pi^{2}(2\ln{L}-1)I_{4}-\frac{\pi^{2}}{2}\gamma_{\mu}\hat{p}\gamma_{\mu}
 -m\, i\, \pi^{2}\big(\ln B(p)\big)I_{4}\big]
\label{4.13}\end{equation}
for $\mu=1,2,3$ and
\begin{equation}J_{4}(L)=-\frac{1}{(2\pi)^4}\big[m\, i\, \pi^{2}(2\ln{L}-1)I_{4}+\frac{\pi^{2}}{2}\gamma_{4}\hat{p}\gamma_{4}
 -m\, i\, \pi^{2}\big(\ln(B(p)\big)I_{4}\big],
\label{4.14}\end{equation}
where
\begin{equation}\ln(B(p))=\int_{0}^{1}\ln(\ell-p^{2}u^{2})du.
\label{4.15}\end{equation}
\begin{Rk}\label{Remark 4.2} The right-hand sides of the equalities \eqref{4.13} and  \eqref{4.14} tend to infinity when $L{\to}\infty$. In this way  the ultraviolet divergence problems
appear.
 \end{Rk}
In order to deal with the considered case, we define  the deviation factor $W_{0}(L,\mu)$  by the formulas
\begin{equation}W_{0}(L,\mu)=(L^{2}/B(p))^{i\phi}\exp(-i\psi)I_{4}\qquad \left(\phi=\frac{me^2}{16\pi^2},\quad \psi=\frac{me^2}{16\pi^2}\right)\label{4.16}\end{equation}
for $\mu=1,2,3$, and $W_{0}(L,4)=-W_{0}(L,1)$.

\emph{Next, we
 use a new representation of $d(p,L)$}: 
\begin{equation}d(p,L)=W_{0}(L,\mu)\tilde{d}(p,L).
\label{4.17}\end{equation}
In view of \eqref{4.16} and \eqref{4.17} we obtain the power series representation of $\tilde{d}(q,L)$:
\begin{equation}\tilde{d}(q,L)=1+e^{2}\tilde{a}_{2}(p,L)+...,
\label{4.18}\end{equation}
It follows that {\it the second  term}  $\tilde{a}_{2}(p,L)$  {\it of series \eqref{4.18} is regular and has the form}
\begin{equation}\tilde{a}_{2}(p,L)=-\frac{1}{32(\pi)^2}\gamma_{\mu}\hat{p}\gamma_{\mu},\quad (1{\leq}\mu{\leq}4).
\label{4.19}\end{equation}
\begin{Rk}\label{Remark 4.3}The  function  $W_{0}(L,\mu)$ given by \eqref{4.16} is a unitary matrix and
satisfies the conditions of  Remark 3.5.   Thus,  $W_{0}(L,\mu)$ is, indeed, the deviation factor.\end{Rk}
\begin{Rk}\label{Remark 4.4}
We have chosen the multiplier $C(p)=\exp(-i\psi)$ (see Remark 3.6) in such a way that $\tilde{a}_{2}(p,L)=0$ when $\hat{p}=0$.\end{Rk}
\begin{Ee}\label{Example 4.5} Let us consider relation \eqref{4.5}, where d(p,L) is a scalar. In this way we  evaluate the second order photon 
self-energy part  \cite[Section 47.3]{AB}.We use some calculations from  \cite{AB} but present an
original interpretation of the results.
\end{Ee} 
The second term $a_{2}(p,L)$ has the form  \cite[Section 47.3]{AB}:
\begin{equation}a_{2}(p,L)=P^{(2)}_{\mu\nu}(p,L)=\frac{1}{(2\pi)^4}\, \tr \,\int_{Q}\frac{\gamma_{\mu}(i\hat{k}-m)\gamma_{\nu}[i(\hat{k}-\hat{p})-m]}{(k^2+m^2)[(p-k)^2+m^2]}dk,
\label{4.20}\end{equation}
where $\tr$ means trace (of a matrix).
    The function $P^{(2)}(p,L)$ is defined by the relation
\begin{equation}P^{(2)}(p,L)=\frac{1}{3}P^{2}_{\mu\mu}(p,L)-\frac{4p_{\mu}p_{\nu}}{3p^{2}}P^{(2)}_{\mu\nu}(p,L).
\label{4.21}\end{equation}
The following asymptotics was obtained   for $P^{(2)}(p,L)$ in  \cite[Section 47.3]{AB}:
\begin{equation}P^{(2)}(p,L)=\frac{8\pi^{2}ie^{2}}{(2\pi)^4}
[-\frac{p^2}{6}(\ln(L^2/m^2)-\frac{5}{6})+p^{2}b]+o(1),
\quad L{\to}\infty,
\label{4.22}\end{equation}
where
\begin{equation}b=\int_{0}^{1}x(1-x)\ln[1+(p^2/m^2)x(1-x)]dx,\label{4.23}\end{equation}
It follows from \eqref{4.22} and \eqref{4.23} that:\\
1) the deviation factor $W_{0}(L)$ is defined by the formula
\begin{equation}W_{0}(L)=(L/m)^{i\phi}\exp(i\psi),\quad \phi=-\frac{e^{2}p^2}{6\pi^{2}},\quad
\psi=\frac{5e^{2}p^2}{2(6\pi)^{2}}; \label{4.24}\end{equation}
2)  the second  term  $\tilde{a}_{2}(p,L)$  of relation \eqref{4.5} is regular and has the form
\begin{equation}\tilde{a}_{2}(p,L)=i\frac{p^2}{2\pi^{2}}b.\label{4.25}\end{equation}
\begin{Rk}\label{Remark 4.6}The function
 $W_{0}(L)$ given by \eqref{4.24} satisfies the conditions of Definition \ref{Definition 3.3}.
 Thus,  $W_{0}(L)$ is, indeed, the deviation factor.
 \end{Rk}
\begin{Rk}\label{Remark 4.7}   It seems interesting that  the expressions of the type $aL^2 \, (a{\ne}0)$
may be found in the intermediate calculations in \cite[Section 47.3]{AB}.
However, in the final result \eqref{4.22} these expressions disappear.\end{Rk}
\begin{Rk}\label{Remark 4.8}
We have chosen the multiplier $C(p)=\exp(i\psi)$ (see Remark 6.3) in such a way that $\tilde{a}_{2}(p,L)$ and
 $\frac{\partial}{\partial{p}}\tilde{a}_{2}(p,L)$  are equal to zero, when $p=0$.\end{Rk}
\begin{Ee}\label{Example 4.9} In order to evaluate the third order vertex part in the case of the
external electron lines, we consider relation \eqref{3.1} \cite[Section 47.4]{AB}.
\end{Ee}
The second term $a_{2}(q,L)$ in  \eqref{3.1} is connected with  the Feynman integral (see  formula (47.50) in \cite[Section 47.4]{AB}):
\begin{equation} a_{2}(q,L)=\Delta_{\mu}^{(3)}(p_1,p_2,q)=-\gamma_{\mu}\frac{e^2}{(2\pi)^2}[-\frac{1}{4}\ln\frac{L^2}{m^2}+
\ln\frac{m}{\lambda}+O(1)],\label{4.23'}\end{equation}
where  $L{\to}\infty,\quad \lambda{\to}0.$
We note that $\lambda$ is the photon ``mass". In the Example 4.9 we have ultraviolet divergence
 ($L{\to}\infty$) and infrared divergence ($\lambda{\to}0$).
It follows from \eqref{4.23'} that
 the deviation factor $W_{0}(L,\lambda)$ is defined by the formula
\begin{equation}W_{0}(L,\lambda)=\exp\{-\gamma_{\mu}\frac{e^2}{(2\pi)^2}[-\frac{1}{4}\ln\frac{L^2}{m^2}+
\ln\frac{m}{\lambda}]\}.\label{4.24'}\end{equation}
\emph{In this way the  ultraviolet divergence
 (when $L{\to}\infty$) and the infrared divergence (when $\lambda{\to}0$) disappear  simultaneously.}
We do not write down the formula for the second term $\tilde{a}_{2}(q,L)$ in \eqref{3.5} since it is rather complicated (see \cite[formula (47.52)]{AB}). Our main goal is the construction and investigation of the deviation factor $W_{0}(L,\lambda)$.
We recall that according to \eqref{4.4} we have
\begin{equation}\gamma_{\mu}=-\gamma_{\mu}^*,\quad \mu=1,2,3.\label{4.25'}\end{equation}
\begin{Rk}\label{Remark 4.10} If $\mu=1,2 \, {\mathrm{or}} \, 3$,
then 
$W_{0}(L,\lambda)$  given by \eqref{4.24'} satisfies the conditions of Definition \ref{Definition 3.3} and Remark \ref{Remark 3.5}. Hence, this $W_{0}(L,\lambda)$   is the deviation factor for the
problem described by \eqref{4.23'}.\end{Rk}
\begin{Rk}\label{Remark 4.11}
We note that $\gamma_4=\gamma_{4}^*$.  Hence, in this case the  matrix $W_{0}(L,\lambda)$ is not unitary for $\mu =4$. It could mean that the corresponding scattering problem  has
no physical sense.\end{Rk}
\section{Dimensional regularization}
An important method for removing divergences is based \cite{COL, POK}  on the transition to the dimension $4-i\varepsilon,\, \varepsilon{\to}+0$. In this section, we plan to show that our approach is  useful
for such problems too. In addition, we compare the two types of divergences: $L{\to}\infty$ and $\varepsilon{\to}+0$.\\
Let us consider the relation
\begin{equation} d(s,\varepsilon)=1+a_{2}(s,\varepsilon)\lambda^{2}+o(\varepsilon),
\label{5.1}\end{equation}
where $s=(p_1+p_2)^2.$
  The corresponding element $d(s)$ of the scattering matrix is defined by the relation
\begin{equation}d(s)=\lim{d(s,\varepsilon)},\quad \varepsilon{\to}0.\label{5.2}\end{equation}
We shall consider the case when the limit on the right-hand side of the equation \eqref{5.2} does not exist.
\begin{Ee}\label{Example 5.1} The following Feynman
integral is used in \cite{COL, POK}:
\begin{equation}a_{2}(s,\varepsilon)=\frac{1}{2}(\mu^{\varepsilon})^2\int_{0}^{1}dx\int\frac{d^{4}k}{(2\pi)^{4}}\frac{1}{[k^2+sx(1-x)-m^2+i\varepsilon)]^{2}}.
\label{5.3}\end{equation}\end{Ee}
It is shown in \cite[p. 104]{POK} that  :
\begin{equation}a_2(s,\varepsilon)=i\frac{1}{16\pi^2\varepsilon}-i\frac{1}{2(4\pi)^2}B_{1}(s)+O(\varepsilon),
\label{5.4}\end{equation}
where
\begin{equation}B_{1}(s)=\gamma-\ln(4\pi)+\int_{0}^{1}dx\ln\frac{m^2-x(1-x)s}{\mu^2}.\label{5.5}\end{equation}
Here $\gamma$ is Euler constant.
Formula \eqref{5.4} implies that $a_2(s,\varepsilon)$ tends to infinity when $\varepsilon{\to}0.$
To obtain formulas without divergences  we use again the notion of the deviation factor.
\begin{Dn}\label{Definition 5.2} The function $W_{0}(\varepsilon)$ is the deviation factor
for  the case described by \eqref{5.1} if the following conditions are fulfilled:\\
1) $|W_{0}(\varepsilon)|=1;$ \\
2) the second term
$\tilde{a}_{2}(s,\varepsilon)$
in the relation
\begin{equation}\tilde{d}(s,\varepsilon)=W_{0}^{-1}(\varepsilon)d(s,\varepsilon)=1+\tilde{a}_{2}(s,\varepsilon)\lambda^{2}+o(\varepsilon)
\label{5.6}\end{equation}
is regular, that is it  converges when $\varepsilon{\to}0.$
\end{Dn}
Relations  \eqref{5.4} and \eqref{5.5} imply that:\\
1) the deviation factor $W_{0}(\varepsilon)$ is defined by the formula
\begin{equation}W_{0}(\varepsilon)=\exp\left({i\frac{\lambda^2}{16\pi^2}\frac{1}{\varepsilon}} \right);\label{5.7}\end{equation}
2) the second term
$\tilde{a}_{2}(s,\varepsilon)$
in the relation \eqref{5.6} has the form
\begin{equation}\tilde{a}_{2}(s,\varepsilon)=-\frac{i}{2(4\pi)^2}B_{1}(s).\label{5.8}\end{equation}

Let us consider another method (see section 4) of regularizing the integral \eqref{5.3}.
We write the corresponding integral
\begin{equation}a_{2}(s,L)=\frac{1}{2}\int_{0}^{1}dx\int_{Q}\frac{d^{4}k}{(2\pi)^{4}}\frac{1}{[k^2+sx(1-x)-m^2]^{2}}.
\label{5.9}\end{equation}
Using \eqref{4.8} we obtain
\begin{equation}a_{2}(s,L)=\frac{i\pi^{2}}{(2\pi)^{4}}\int_{0}^{1}[2\ln{L}-\ln(m^2-x(1-x)s)-1]dx+o(1),
\quad L{\to}\infty.\label{5.10}\end{equation}
The relations \eqref{5.9} and \eqref{5.10} imply that:\\
1) The deviation factor $W_{0}(L)$ is defined by the formula
\begin{equation}W_{0}(L)=L^{i\phi},\quad \phi=\frac{\lambda^2}{16\pi^{2}} \label{5.11}\end{equation}
2) The second term
$\tilde{a}_{2}(s,L)$  has the form
\begin{equation}\tilde{a}_{2}(s,L)=-\frac{i}{2(4\pi)^2}B_{2}(s).\label{5.12}\end{equation}
where
\begin{equation}B_{2}(s)=\int_{0}^{1}\ln(m^ 2-x(1-x)s)dx-1\label{5.13}\end{equation}
\begin{Rk}\label{Remark 5.3} The function $W_{0}(L)$ given by \eqref{5.11}    satisfies all the conditions of Definition \ref{Definition 3.3}
and is, indeed, the deviation factor.
\end{Rk}
\begin{Rk}\label{Remark 5.4} Formulas \eqref{5.5},\eqref{5.13} and \eqref{5.8}, \eqref{5.12} imply that:
$$\tilde{a}_{2}(s,\varepsilon)-\tilde{a}_{2}(s,L)\equiv{\mathrm{const}}.$$
  Since constants (as above) are not essential in the scattering theory, one can say that both methods of regularization (i.e., the cases $\varepsilon{\to}0$ and $L{\to}\infty$) bring the same result.\end{Rk}
\section{Conclusion}
1. In this paper, we presented  a new approach to the divergence problems in QED.
We   started with  such important examples as
the radial  Schr\"{o}dinger and  Dirac equations with Coulomb-type potentials.
For these cases, we studied the divergence problem in a mathematically   rigorous way.
For the Dirac equation,  the obtained results are new.

2.  Further in the paper we investigated  the famous Feynman integrals. In Example 4.9, we used some calculations from the book \cite{AB}, but presented our
original interpretation of the results.

3. Our approach can be applied to various other cases. In particular, many interesting examples from \cite{AB}
may be interpreted  similar to  the way the Example 4.9 is treated in the present paper.
We show also that our approach can be used in the 4-$\varepsilon$ dimensional case (see Section 5).

4. It is interesting that two methods of removing the divergences ($\varepsilon{\to}0$  and $L{\to}\infty$) produce the same results (see Example 5.1).
This property is an analog of the ergodic property,
when stationary and dynamical scattering operators coincide (see \cite{Sakh5, Sakh4}).

\vspace{1em}

\noindent{\bf Acknowledgment.}
 {The author is grateful   to  A.L. Sakhnovich  for his help in the preparation of the manuscript.}

\appendix

\section{Appendix. Generalized wave operators}
1. Wave operators play an essential role in many problems of mathematical physics (see \cite{RS}).
However, the wave operators do not exist, when the initial and (or) final states of the system cannot be regarded as free.
In these cases, one has to consider the {\it generalized wave operators}  (see, e.g.  \cite{Dol, Sakh2, Sakh3, Sakh10}).

Below we introduce the notions  of the generalized wave operators $W_{\pm}(A,A_{0})$  \cite{BuM, Sakh2}  and of the deviation factor $W_{0}(t)$ \cite{Sakh5,Sakh4},
where $A$ and $A_0$ are linear self-adjoint (not necessary bounded) operators acting in some Hilbert space $H$. The orthogonal projector on the
absolutely
continuous subspace  $G_0$ of $A_0$ is denoted by $P_0$.
 \begin{Dn}\label{Definition 7.1}
 Operator functions $W_{\pm}(A,A_{0})$ and  $W_{0}(t)$ are called the generalized wave operators and a deviation factor, respectively, if
the following conditions are fulfilled:

1. The limits
\begin{equation}W_{\pm}(A,A_{0})
=\lim_{t{\to}\pm\infty}[e^{iAt}e^{-iA_{0}t}W_{0}^{-1}(t)]P_0
\label{7.1}\end{equation}
exist in the sense of strong convergence. 

2. The operators $W_{0}(t)$ and  $W_{0}^{-1}(t)$ are acting in the subspace $G_0$,
are
unitary for all $t$ and
\begin{equation}\lim_{t{\to}\pm\infty}W_{0}(t+\tau)W_{0}^{-1}(t)P_0=P_0,\quad \tau=\overline{\tau}.
\label{7.2}\end{equation}

3. The following commutation relations  hold for arbitrary values $t$ and $\tau$:
\begin{equation}W_{0}(t)A_{0}P_0=A_{0}W_{0}(t)P_0,\quad
W_{0}(t)W_{0}(t+\tau)P_0=W_{0}(t+\tau)W_{0}(t)P_0.
\label{7.3}\end{equation}
\end{Dn}
If  $W_{0}(t)=P_0$, then the operators $W_{\pm}(A,A_{0})$ are usual wave
operators.
\begin{Dn}\label{Definition 7.2}The generalized scattering operator $S(A,A_0)$
is defined by the formula:
\begin{equation}S(A,A_0)=W^{*}_{+}(A,A_0)W_{-}(A,A_0),\label{7.4}
\end{equation}
where $W_{\pm}(A,A_{0})$ are generalized wave operators.\end{Dn}
The operator $S(A,A_0)$ is a unitary mapping of $G_0$  onto itself and
\begin{equation}A_{0}S(A,A_0)P_{0}=S(A,A_0)A_{0}P_{0}.\label{7.5}\end{equation}
2. Let us consider the equation
\begin{equation}i\frac{d\psi}{dt}=A\psi,\quad A=A_{0}+{\varepsilon}A_1, \label{7.6}\end{equation}
where $A$  and $A_{0}$ are self-adjoint operators, and $A_1$ is  a self-adjoint perturbation operator.  The solution of \eqref{7.6} has the form
\begin{equation}\psi(t)=e^{-iAt}\psi(0).\label{7.7}\end{equation}
If the generalized wave operators $W_{\pm}(A,A_0)$ exist , then
\begin{equation}\psi(t)=e^{-iAt}\psi(0){\sim}e^{-iA_{0}t}\psi_{\pm},\quad t{\to}\pm\infty,
\label{7.8}\end{equation}
where
\begin{equation}\psi_{0}=W_{\pm}(A,A_0)\psi_{\pm},\quad \psi_{+}=S(A,A_0)\psi_{-}.\label{7.9}
\end{equation}
\begin{Rk}\label{Remark 7.3} The deviation factors are not uniquely defined. If $W_{0}(t)$ is the deviation factor and $C$ is a unitary operator,
then $CW_{0}(t)$ is  the deviation factor too.
The choice of multipliers $C$ depends on the particular physical problem
under consideration.\end{Rk}

\end{document}